\documentclass[12]{article}
\usepackage{amsmath,amssymb,amsthm}
\usepackage{mathtools}
\usepackage{mathrsfs}
\usepackage{bm}
\usepackage{slashed} %FeynmanSlashNotation
\usepackage{multirow}
\usepackage{tikz}
\usepackage[utf8]{inputenc}
\usepackage{fullpage}
\usepackage{arydshln}
\usepackage{setspace}
\usepackage[margin=1in]{geometry}
\usepackage{dsfont}
\usepackage{graphicx}
\usepackage{cite}
\graphicspath{ {graphs/} }

\newcommand{\be}{\begin{eqnarray}}
\newcommand{\ee}{\end{eqnarray}}

\usepackage{hyperref}
\hypersetup{colorlinks=true,linkcolor=blue,anchorcolor=blue,citecolor=green,filecolor=blue
urlcolor=blue,bookmarksnumbered=true,pdfview=FitB}

\usepackage{ulem}

\begin{document}
\title{\textbf{Critical Coupling for Two-dimensional $\phi^4$ Theory in Discretized Light-Cone Quantization}}
\author{James P. Vary$^1$, Mengyao Huang$^1$, Shreeram Jawadekar$^1$, Mamoon Sharaf$^1$,\strut\\ 
Avaroth Harindranath and Dipankar Chakrabarti$^2$}
\maketitle
\begin{center}
\textit{$^1$Department of Physics and Astronomy, Iowa State University, Ames, Iowa, U.S.A., 50011} \\
\textit{$^2$Department of Physics, Indian Institute of Technology Kanpur, Kanpur, India, 208016} 
\end{center}

\begin{abstract}
We solve for the critical coupling in the symmetric phase of two-dimensional $\phi^4$ field theory 
using Discretized Light-Cone Quantization.  We adopt periodic boundary conditions, neglect the
zero mode, and obtain a critical coupling consistent with the critical coupling reported using conformal truncation in light-front 
quantization.  We find a 17\% difference from the critical coupling reported with light-front 
quantization in a symmetric polynomial basis.
\end{abstract}

\section{Introduction}

%Introduction to what we do in this work and how we distinguish our approach from that of others.
% * * * * * * * 
% Following adapted from Hari and Dipankar's file: phi42-sym1-note1.tex received on March 6, 2021

It is  well known  that  most of our intuition about spectra 
and wave functions in quantum mechanics come from solving Hamiltonians and 
 a variety of methods exist for this purpose. 
In molecular,  atomic and subatomic quantum systems, sophisticated and efficient numerical
approximation techniques have been developed and refined over the years to solve 
quantum many-body Hamiltonians. One would like to make 
use of the intuition developed and experience gained in solving 
non-relativistic systems to  devise
methods to solve systems governed by relativistic quantum field 
theories. But Hamiltonian methods were rarely developed for this purpose.
{\textquotedblleft}This is a pity{\textquotedblright}, as was noted 
sometime back \cite{kgw65}. 
There are major stumbling blocks to this path caused by the presence of 
infinitely many degrees of freedom and the mandatory need for 
renormalization. Especially noteworthy is the severe 
divergence caused by vacuum
processes which are addressed analytically in the perturbative 
framework. One has to learn how to handle them or at least
how to side step them  in the non-perturbative Hamiltonian framework.      

It was Dirac \cite{dirac} who first pointed out  the advantages of 
light-front dynamics among different Hamiltonian formulations of relativistic 
many-body theories. 
Great interest in light-front dynamics arose because light-front 
quantization provides the theoretical basis for the celebrated parton 
model \cite{feynman, bjorkenpaschos} which facilitates an 
intuitive understanding of high energy processes.
Many additional virtues of light-front quantization have been 
identified and studied since the work of Dirac, but a 
non-perturbative numerical approach was missing. 
Discretized Light-Cone Quantization (DLCQ) 
\cite{pb85, earlydlcq,pr,Hiller:2016itl}
was proposed in the mid 1980s 
as a novel, non-perturbative numerical technique to solve quantum field 
theories. 

Two-dimensional interacting  scalar field ($\phi^4$)  theory, in spite of 
being the simplest of
interacting quantum field theories, has a rich structure and has been the
subject of both rigorous mathematical analysis \cite{bs,gj} and various 
non-perturbative 
numerical approaches. Not surprisingly, soon after the proposal of DLCQ, 
 it was applied to
the symmetric phase of two-dimensional $\phi^4$  theory\cite{hv87}. As a result, 
many strengths and a few weaknesses of the 
method  were identified already in those early days. It was known that the 
system 
undergoes a phase transition from the symmetric phase to the symmetry broken 
phase at strong coupling as a result of quantum fluctuations.  Among the 
observables calculated in Ref. \cite{hv87} was the critical coupling for 
the vanishing mass gap in the odd particle sector as a function of the 
dimensionless total longitudinal momentum  $K$. It was realized that 
the continuum limit is reached by taking $K \rightarrow \infty$ which
requires reliable extrapolation of  finite $K$ results. 
An extrapolation 
was attempted in Ref. \cite{hv88} using finite $K$ results with maximum 
$K$ of 20. The limited data available was extrapolated to $K$ = 100 and the
corresponding critical coupling was found to be $\approx$ 33.

The critical coupling has been calculated since then, with various
nonperturbative techniques within both
Lagrangian and Hamiltonian frameworks. Within the light-front Hamiltonian 
formalism,  two works have recently emerged which are of immediate
relevance to DLCQ, namely Burkardt {\em et al.} 
\cite{bch} and Anand {\em et al.} \cite{agkkw}.           

% ********************************************************************:
In Burkardt {\em et al.},  the Hamiltonian is 
expressed in the Fock space basis 
in the continuum. Here, using the positivity of the light-front longitudinal
momentum, the terms in the Hamiltonian containing  pure creation 
or annihilation operators are dropped. The  state also is expanded in a Fock 
basis where the coefficients are multi-particle wave functions.
Then  the  multi-particle wave functions
are expanded in terms of  symmetric polynomials.
These polynomials span the entire range of the momentum fraction $x$ 
from 0 to 1, {\em including} the end points.  A finite dimensional
Hamiltonian is obtained by truncating the maximum number of bosons
and the maximum number of basis functions used.

In Anand {\em et al.}, the light-cone conformal truncation is used. 
As explained in detail in Ref. \cite{Anand:2020gnn}, 
one starts from the conformal limit of the two-dimensional $\phi^4$ theory, 
which is a free massless theory. 
A complete set of basis states of conformal quadratic Casimir {\cal C} is constructed using primary operators.
These states are used to express
the Hamiltonian of the two-dimensional $\phi^4$ theory in light-cone 
quantization.
After truncating the basis
to states with Casimir eigenvalue  ${\cal C}$  below some 
threshold ${\cal C}_{\rm max}$, 
 the resulting finite dimensional Hamiltonian matrix is numerically  
diagonalized. The positivity of the 
light-front longitudinal momentum is employed in this work.

By the use of basis functions in the continuum in the above two 
references, sensitivity to the small longitudinal momentum region can be
explored  and handled very efficiently. On the other hand in the previous
and current work on two-dimensional $\phi^4$ theory with Periodic Boundary 
Conditions (PBC), exactly zero longitudinal momentum, which is a constrained 
(non-dynamical) mode in the symmetric phase, is dropped. Furthermore, 
DLCQ, in contrast with Refs~\cite{bch,agkkw}, uses a uniform grid. 
Therefore, an arbitrarily small longitudinal
momentum can be accessed only in the continuum limit. Thus, by approaching
the continuum limit in DLCQ numerically and comparing the resulting value 
for the critical coupling with those of Refs~\cite{bch,agkkw}, 
we can quantify the efficiency and legitimacy of the DLCQ program.       
% ******************************************************************

Motivated by the work of Rozowsky and 
Thorn \cite{rt}, who argued that the exactly zero longitudinal momentum mode
(hereafter referred to as the zero mode) is not necessary to describe 
spontaneous breaking of symmetry, the symmetry broken phase of
two-dimensional $\phi^4$ theory was investigated in DLCQ in detail with both 
PBC with the omission of the zero mode 
\cite{chmpv} and Anti-Periodic Boundary Conditions (APBC) \cite{chmv, chv} 
where the zero mode is naturally absent. 
Masses of low-lying states, their parton distributions and coordinate space
($x^-$) profiles were successfully calculated. These accumulated results 
provide another motivation to revisit the symmetric  phase  of two-dimensional $\phi^4$ theory in DLCQ.

To  analyze realistic quantum field theory problems in 3+1 dimensions,
currently, the most efficient light-front Hamiltonian approach is 
the Basis Light-Front Quantization (BLFQ)  method \cite{Vary:2009gt,Zhao:2020gtf}, which 
frequently employs DLCQ in the light-front longitudinal direction and orthonormal 
basis functions in the transverse space.   
BLFQ utilizes the vast  practical experience gained in solving strongly 
interacting nuclear many-body systems and employs the same methodology 
and techniques to diagonalize the Hamiltonians of enormous dimensionality.
Thus the lessons learned in studying the convergence and extrapolation
issues of DLCQ applied to strong coupling problems in 1+1 dimensions are 
also directly relevant for the BLFQ program.         
  
Within the Hamiltonian-based approaches, in order to quantitatively assess
the strengths and weaknesses of DLCQ, corresponding
calculations in the Instant form with a Fock-space based Hamiltonian are
a must. This was lacking for a long time. Fortunately, results of
such calculations in both the symmetric \cite{rv1} and broken 
\cite{rv2} phases of two-dimensional $\phi^4$ theory
have become recently available. 
We also note that, over the past two decades there 
have also been many studies of the critical coupling
of two-dimensional $\phi^4$ theory. See,  for example, 
Refs. \cite{lsl,ts,sl,mho,bpg,ssv,dt}. However, with the exception of 
Refs. \cite{bch} and \cite{agkkw}, 
a direct comparison of the value of critical coupling via DLCQ presented
in this work with the other works cited is not possible because of the 
matching problem between equal-time and light-front 
methods \cite{bch,  fkw}.  

Our primary focus in the present work is to obtain the critical coupling of two-dimensional $\phi^4$ theory 
in the continuum limit of DLCQ for the transition from the symmetric phase to the broken 
phase.  To be more precise, we extract three separate critical couplings that each produce a vanishing
mass gap at finite resolution, $K$. These are the critical couplings for the lowest state 
of the odd sector, the lowest state of the even sector and the first
excited state of the odd sector.  We find agreement of all three critical couplings in the continuum 
limit ($K \rightarrow \infty$), implying degeneracy at a vanishing mass gap, to within numerical uncertainties, which we interpret 
as the signal for spontaneous symmetry breaking. 

% * * * * * * *

\section{Theoretical Framework}
\subsection{Two-dimensional $\phi^4$ theory in Discretized Light-Cone Quantization}

We seek to solve a Hamiltonian ($H$) eigenvalue problem expressed in a suitable
basis to obtain the low-lying mass spectroscopy.  In this work, we solve for and present 
the eigenvalues of the mass-squared operator proportional to $H$.
In principle, once the eigenvectors are obtained,  we could evaluate matrix elements 
of additional observables $O$ but we defer those efforts to a future project.
For the present work, we aim to
obtain that spectroscopy in the region of critical coupling producing a vanishing
mass gap at a sequence of basis space cutoffs $K$. We then perform detailed extrapolations of these
results to the continuum limit (infinite matrix dimension) accompanied by an uncertainty analysis.

The Hamiltonian for two-dimensional $\phi^4$ theory in DLCQ 
was first presented in Ref.~\cite{hv87}.  
We start from the two-dimensional Lagrangian density
\be
%{\cal L} = \frac{1}{2} \partial^\mu \phi \partial_\mu \phi -
{\cal L} =   \frac{1}{2}\partial^+ \phi \partial^- \phi -
 \frac{1}{2} \mu^2 \phi^2 -  \frac{\lambda}{4!} \phi^4
\ee
with the light-front variables defined by $ x^\pm = x^0 \pm x^1$.

The Hamiltonian density
\be
{\cal P}^- =
  \frac{1}{2} \mu^2 \phi^2 +  \frac{\lambda}{4!} \phi^4
\ee
defines the Hamiltonian
\be
P^- & = & \int dx^- {\cal P}^-~
     \equiv  \frac{L}{2 \pi} H,
\ee
where $L$ defines our compact domain $ - L \le x^- \le +L$. Throughout this
work we address the energy spectrum of $H$.

The longitudinal momentum operator is
\be
P^+ & = & \frac{1}{2} \int_{-L}^{+L} dx^- \partial^+ \phi \partial^+ \phi
 \equiv  \frac{2 \pi}{L}K
\ee
where $K$ is the dimensionless longitudinal momentum operator. The mass-squared 
operator $M^2 = P^+ P^- = KH$ whose eigenvalues we present in this work.

% ************************************************************************
With PBC  we can write
\begin{equation}
\phi= \phi_0 + \Phi,
\end{equation}
where $\phi_0$ is the zero mode operator and $\Phi$ contains non-zero modes. 

By integrating the equation of motion  
$\partial^+\partial^- \phi +\mu^2
\phi +  \frac{\lambda}{3!}\phi^3 =0~$
over the longitudinal space, 
one can show that (1)  the zero mode operator vanishes in the free field theory,
and (2) in the interacting theory, the zero mode operator is constrained and 
obeys a nonlinear operator equation. 
Thus incorporating the zero mode in
DLCQ is a nontrivial problem. To the best of our knowledge, this problem 
has not yet been solved in a satisfactory manner. When the constraint 
equation is solved perturbatively, it
leads to additional interaction terms in the Hamiltonian. 
For some processes studied, contributions of these new terms, however
were found to vanish \cite{hswz, hmv} in
the infinite volume  limit ($L \rightarrow \infty$). 
 
For the non-zero mode part $\Phi$ we use the solution of the equation of
motion for the free massive scalar field in the continuum theory, which 
provides a convenient Fock space basis: 
\begin{equation}
\Phi(x^-) = \int {dk^+    \over 2 (2 \pi) k^+} \theta(k^+) \big [
a(k^+)e^{-\frac{i}{2}k^+ x^-} + a^{\dagger}(k^+) e^{\frac{i}{2}k^+ x^-} \big ]~.
\end{equation} 
Note that zero mode is absent in this expression.
In DLCQ, the corresponding field expansion is 
% ************************************************************************
\be
\Phi(x^-) = \frac{1}{\sqrt{4 \pi}} \sum_n \frac{1}{\sqrt{n}}
\left [a_n e^{-i \frac{n \pi}{L} x^-} + a_n^\dagger e^{i \frac{n
\pi}{L} x^-} \right ],
\ee
with $ n = 1, 2, 3,\ldots$ with PBC and $n = \frac{1}{2}, \frac{3}{2}, \frac{5}{2}\ldots$ with APBC.
% (APBC): with $ n = \frac{1}{2}, \frac{3}{2},\ldots. $

% * * * * * * * Above is work area with inserts from phionset_v4.tex

     The normal ordered dimensionless longitudinal momentum operator 
\begin{equation}
 K~  = ~ \sum_n~ n a_n^\dagger a_n.
 \end{equation}

The normal ordered Hamiltonian  is given by

\be
H & = & \mu^2 \sum_n \frac{1}{n} a_n^\dagger a_n
+ \frac{\lambda}{4 \pi} \sum_{k \le l, m\le n}~ \frac{1}{N_{kl}^2}
~ \frac{1}{N_{mn}^2}~
\frac{1}{\sqrt{klmn}}
 a_k^\dagger a_l^\dagger a_n a_m
\delta_{k+l, m+n} \nonumber \\&~& + \frac{\lambda}{4 \pi} \sum_{k, l \le
m\le n}~ \frac{1}{N_{lmn}^2}~ 
\frac{1}{\sqrt{klmn}}~
  \left [
a_k^\dagger a_l a_m a_n + a_n^\dagger a_m^\dagger a_l^\dagger a_k \right ]~
\delta_{k, l+m+n}
\ee
with
\be N_{lmn} & = & 1 ,~ l \ne m \ne n, \nonumber \\
            & = & \sqrt{2!},~ l=m \ne n, ~ l \ne m=n,\nonumber \\
            & = & \sqrt{3!},~ l=m=n,
\ee
and
\be
N_{kl} & = & 1, k \ne l,~ \nonumber \\
       & = & \sqrt{2!},~ k=l.
\ee

\subsection{Symmetries and Vanishing Mass Gaps}

Since the Hamiltonian exhibits the $ \phi \rightarrow - \phi$ symmetry, the
even and odd particle sectors of the theory are decoupled. In the ``symmetric
phase" of the theory (positive bare mass-squared $\mu^2$) which we investigate
here, the solutions at weak positive values of $\lambda$ have simple structure.
In the odd sector, the lowest solution is dominated by a single boson carrying
all the light-front momentum $K$.  In the even sector, the lowest solution is dominated
by two bosons each carrying $K/2$.

We do not invoke mass renormalization so that, with increasing coupling, these two lowest 
states will each decrease towards zero mass.  A separate value of the critical coupling at fixed K 
is obtained when each of these mass gaps vanishes.  Mass gaps for higher states in each
sector can also vanish at fixed K but at successively larger values of the critical coupling.  
In line with expectations, we find that, as a function of increased coupling, the lowest state of the odd sector 
vanishes before the lowest state of the even sector at each value of $K$. 

For the present work, we are interested in calculating a set of critical couplings that produce 
corresponding vanishing mass gaps over a range of computationally accessible values of $K$.  
With these results, we then carry out extrapolations
of these critical couplings to their continuum limits ($K \rightarrow \infty$).  We perform these
calculations for the two lowest states of the odd sector and the lowest state of the even sector.
We raise and answer the question, to within numerical precision, whether these critical couplings 
are the same or distinct in the continuum limit.  We find that they are the same and we compare our
continuum limit result for the critical coupling with results obtained with other light-front methods~ \cite{bch,agkkw}.

As mentioned above, many interesting questions can be addressed with detailed studies of the 
light-front wave functions at the critical coupling and as a function of $K$.  In addition, one can
examine the sensitivity to choice of boundary conditions, for example, by performing corresponding 
studies with APBC.  We defer these and other valuable topics to 
future research projects.

\subsection{Methods of Solving for the Low-Lying Spectroscopy at Each $K$}

Since the interacting theory is a function of a single dimensionless variable  $\frac{\lambda}{\mu^2}$,
we adopt $\mu^2 = 1$ for convenience.  Thus, our results for functions of  
$\frac{\lambda}{\mu^2}$ are the same as our results for functions of $\lambda$.

Results presented here were obtained using Cori, a supercomputer at the National Energy Research
Computing Center (NERSC)~\cite{NERSC}, 
with the Many Fermion Dynamics (MFD) code adapted to bosons \cite{mfd,chmv,chmpv,chv}. The Lanczos diagonalization 
method is used in a highly scalable algorithm that allows us to proceed to high enough values 
of $K$ for smooth $ K \rightarrow \infty $ extrapolations. 

In order to further assure the reliability of our results, two additional and independent special-purpose codes were written
and employed in these calculations. These two codes are called 
``lfphi4MH-00''~\cite{Mengyao_code} and ``LFHC\_phi4\_SJ''~\cite{Shreeram_code}.
Both codes are MPI parallelized. The code lfphi4MH-00 (LFHC\_phi4\_SJ) is written in C (Fortran90) and employs the diagonalization
package(s) ``DSYEV'' from LAPACK~\cite{lapack99} (``petsc''~\cite{package2a1,package2a2} and ``slepc''~\cite{package2b})). 

While the low-lying eigenvalues provided by MFD were obtained to about
seven significant figures (single-precision with 32-bit words), the eigenvalues from  lfphi4MH-00 and LFHC\_phi4\_SJ
were obtained to 14 significant figures (double precision or 64-bit accuracy).  All three codes produced the same eigenvalues
to within their respective precisions.  The two new codes are not yet optimized to run on supercomputers so their use is
limited in the present work.  Our presented results are obtained from diagonalizations for $K \leq 42$ in double-precision with lfphi4MH-00 and LFHC\_phi4\_SJ
while diagonalization results for $K > 42$ are obtained in single-precision with MFD.

We have performed calculations on meshes of values of $K$ for the even and the odd sectors 
as detailed below. For each sector and each $K$ we perform a set of calculations over
a small range in $\lambda$ sufficient to determine the critical coupling for the vanishing mass gap.
In order to gain an impression of the computational effort, we present the dimensions of $H$ in DLCQ
with periodic boundary conditions at representative values of $K$, both without 
and with boson number truncation in Table~\ref{Table_of_dimensions}.

\begin{table}[ht]
\begin{centering}
\begin{tabular}{|c|c|c|c|c|c|c|}
\cline{2-7} 
\multicolumn{1}{c|}{} & \multicolumn{6}{c|}{Matrix Dimension}\tabularnewline
\hline 
K & Odd sector & Even sector & Odd sector (16) &  Even sector (16)  & Odd sector (8) &  Even sector (8) \tabularnewline
\hline 
\hline 
$\mbox{16}$ & 113  & 118 & 113  & 118 & 87 &  99 \tabularnewline
\hline 
$\mbox{32}$ & 4163 & 4186 & 3774 &  3891 & 1426 &  1893 \tabularnewline
\hline 
$\mbox{48}$ & 73593 & 73680 & 54486 &  58054 & 9616 &  15083 \tabularnewline
\hline 
$\mbox{64}$ & 870677 & 870953 & 488759 &  542632 & 41171 &  75092 \tabularnewline
\hline 
$\mbox{80}$ & 7897846 & 7898630 & 3186613 &  3696386 & 133295 &  278203 \tabularnewline
\hline 
$\mbox{88}$ & 22053415 & 22054694 & 7428056 &  8810476 & 222513 &  494131 \tabularnewline
\hline 
$\mbox{96}$ & 59056128 & 59058176 & 16465206 &  19970504 & 356993 &  840816 \tabularnewline
\hline 
\end{tabular}
\par\end{centering}

\caption{Representative matrix dimensions for scalar $\phi^4$ in DLCQ with periodic boundary conditions
omitting the zero mode. Numbers in parenthesis in column headings signify the maximum number of bosons
allowed which results in a truncated basis.} 
\label{Table_of_dimensions}
\end{table}

It is noteworthy that calculations near a vanishing mass gap involve strong coupling and increased
level density (compared to weak coupling at the same $K$) near the lowest state of the system.  
Both of these features induce the need for considerable care
in attaining and assuring the stability of numerical precision.  One can note, for example, that 
thousands of Lanczos iterations are required for the larger dimensional matrices solved for the 
lowest-lying solutions in the present effort.  As $K$ increases, the number of Lanczos iterations 
increases for eigenvalues near the critical coupling and the numerical precision begins to erode.
Future efforts will focus on improving the efficiency of the double-precision codes in order to achieve greater 
accuracy at higher $K$ and to improve the results for the critical coupling in the continuum limit.

\section{Numerical Results, Polynomial Fits and Extrapolations}

\begin{figure}[ht!]
\centering
\includegraphics[width=15.65cm]{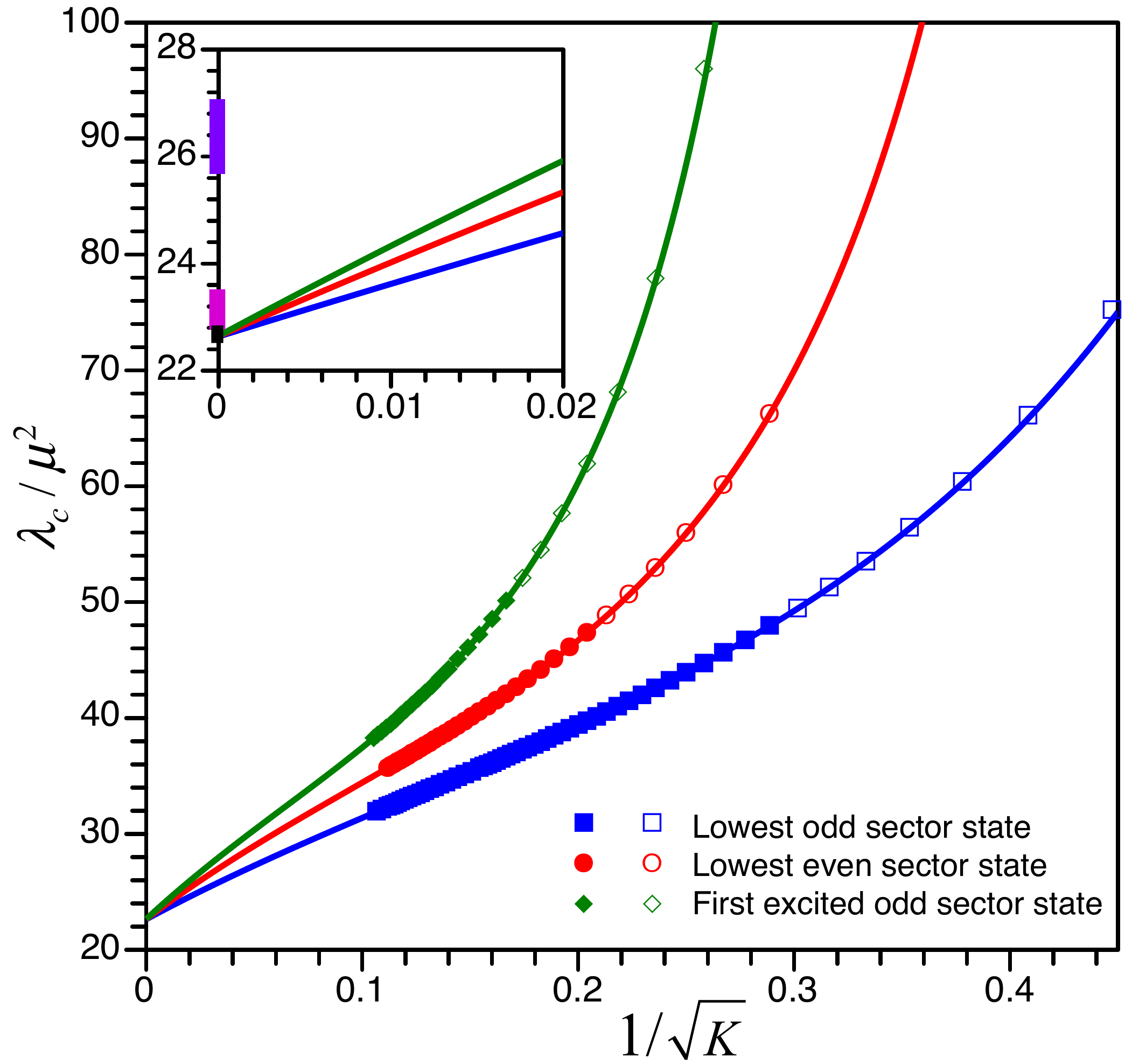}
\caption{Critical couplings $\lambda_c/\mu^2$ for vanishing mass gaps as a function of $1/\sqrt{K}$ 
obtained with DLCQ using periodic boundary conditions.  Solid (open) symbols represent DLCQ results for 3 states, 
according to the legend, that are used (not used) in the fits.  The smooth curves are results for $f_1$ (blue), $f_2$ (red) 
and $f_3$ (green) resulting from procedures using polynomial fits as discussed in the text.  The extrapolated value 
is the intercept on the vertical axis of $22.64\pm 0.17$.
%\textcolor{red}{ $22.64\pm 0.17$ (preliminary)}. 
The inset provides an expanded view of the region approaching the continuum limit.  A vertical black bar represents
our approximate systematic uncertainty which dominates our overall uncertainty (see text).
%The region in the box at the lower left is enlarged with an inset in order to show the continuum limit in more detail. 
The inset also presents, for comparison, a vertical purple bar (magenta bar) on the vertical axis representing the critical coupling result 
of Ref. \cite{bch} (\cite{agkkw}) which is $26.39\pm 0.63$ ($23.12\pm 0.38$). 
%  Directly output from Deltagraph:   Critical_Couplings_SP_page23_Fig1_rev3.pdf 
%  Processed through Powerpoint to revise axis labels: Critical_Couplings_SP_page23_Fig1_rev4.pdf
%\textcolor{red}{Possibly stretch the vertical size of the inset and take the max of the horizontal axis = 0.01}
}
\label{fig:Critical couplings for three states}
\end{figure}

As mentioned above, we focus our attention on the lowest two mass eigenstates of the odd sector and the lowest mass eigenstate
of the even sector.  For these three states we obtain the critical coupling where the eigenvalue vanishes at each $K$.
While details will be introduced and discussed below, we present here a brief overview of our main results.  Fig.~\ref{fig:Critical couplings for three states}, 
displays the critical couplings for the vanishing mass gaps of these three low-lying states as functions of $1/\sqrt{K}$.  The symbols represent calculations 
using DLCQ for each of the three states indicated in the legend and the curves represent polynomial fits using these calculated results 
according to procedures discussed below.

We obtain the continuum limit by taking our functional fits to the limit $K \rightarrow \infty $ as indicated by their
intercepts with the vertical axis in Fig.~\ref{fig:Critical couplings for three states}.  The inset provides an enlarged view of these functions as
they approach the continuum limit. We obtain a result of $22.64\pm 0.17$ 
for the critical coupling for these three states.  
Our result can be compared with the critical coupling result of Ref. \cite{bch} (\cite{agkkw}) which is 26.39 $\pm 0.63$ (23.12 $\pm 0.38$).  Hence, we are consistent with
the result of Ref.~\cite{agkkw} within the respective quoted uncertainties.  Our result is presented as a vertical black band on the vertical axis
of the inset to Fig.~\ref{fig:Critical couplings for three states} while those of Ref.~\cite{bch} and Ref.~\cite{agkkw} appear as vertical purple 
and vertical magenta bands respectively.

\subsection{Lowest state of the odd sector}

We now present details underlying the results displayed in  Fig.~\ref{fig:Critical couplings for three states}.
For the lowest state of the odd sector, we fit a 5th-degree polynomial in $1/\sqrt{K}$ to the DLCQ results for the coupling 
that provides a vanishing mass gap for the lowest mass-squared eigenvalue at each $K$.  We determined our preferred functional form
and variable by testing many functional forms and variables\footnote{For example, we tried various polynomial fits as functions of 
$1/K$ and $1/K^{3/2}$.  We also tried exponential fits as well as fits in the form $A*K^{B}+C$.  We ultimately settled on a polynomial 
as a function of $1/\sqrt{K}$.}.  Our selected function and variable provide robust extrapolations reasonably
independent of the range of DLCQ results included in the fit (see Sec. \ref{sec:R&UQ}).  

To be more specific, we determine the function
\begin{eqnarray}
f_1(z)=& \sum_{j} A_j  z^j, 
\label{Eq:f_1}
\end{eqnarray}
where $z = 1/\sqrt{K}$,  the subscript ``$1$'' specifies the function for the lowest state of the odd sector and $j$ runs from zero to $5$.  
All fits presented in this work are performed by minimizing the mean square deviation
between the adopted function and the DLCQ results. The coefficients of the fits are tabulated in Table \ref{Tab:Expansion Coefficients}. 
We will discuss the residuals of our fits after we introduce the other fits.

\begin{table}[ht]
\begin{centering}
\begin{tabular}{|c|c|c|c|}
\cline{2-4} 
\multicolumn{1}{c|}{} & \multicolumn{3}{c|}{Functions represented by Polynomials }\tabularnewline
\hline 
j & $f_1(1/\sqrt{K})$ & $R_{2/1}(1/\sqrt{K})$ & $R_{3/1}(1/\sqrt{K})$ \tabularnewline
\hline 
\hline 
$\mbox{0}$ & 2.263806133E+1  & 1.000266817 & 1.000463053  \tabularnewline
\hline 
$\mbox{1}$ & 9.991051426E+1 & -- & --   \tabularnewline
\hline 
$\mbox{2}$ & -1.748897600E+2 & -7.877324010E-1 &  7.039552300E-2  \tabularnewline
\hline 
$\mbox{3}$ & 5.409501766E+2   & -- & --  \tabularnewline
\hline 
$\mbox{4}$ & -5.119313071E+2 &  1.569234605E+1 & 7.035735124E+1  \tabularnewline
\hline 
$\mbox{5}$ & 7.891188565E+2   & -- & --    \tabularnewline
\hline 
$\mbox{6}$ & -- & -1.332622505E+2 & -7.658799638E+2  \tabularnewline
\hline 
$\mbox{7}$ & -- & -- & --    \tabularnewline
\hline 
$\mbox{8}$ & -- & 8.179742782E+2 & 9.511647022E+3  \tabularnewline
\hline 
\end{tabular}
\par\end{centering}

\caption{Coefficients of polynomial fits using functions described in the text. Coefficients of functions 
($f_1$,  $R_{2/1}$, $R_{3/1}$) defined in Eqs. (\ref{Eq:f_1}, \ref{Eq:R21aspoly},  \ref{Eq:R31aspoly}) are obtained by least squares fits to DLCQ results over the ranges 
(12-88, 24-80, 36-90) with (52, 29, 19) points respectively. }
\label{Tab:Expansion Coefficients}
\end{table}
% See Deltagraph file "Critical_Couplings_SP..."

In order to limit the role of finite K artifacts that play a larger role at lower values of $K$, 
we selected our 52 DLCQ results in the range $12 \leq K \leq 88$ to determine the coefficients in Eq. \ref{Eq:f_1}.
This range reaches the maximum for which we obtained DLCQ results for the lowest state of the odd sector. 
 In particular, we employed 31 data points with unit increments over $12 \leq K \leq 42$, 19 data points with
increments of two in $K$ over $42 < K \leq 80$,  and the points at $K=84$ and $K=88$.  The choice of
mesh in $K$ is arbitrary but aimed to provide a significant number of points in the range where 
available computational resources were sufficient and a robust fit could be obtained. 
We display the resulting fit 
as a solid blue line in Fig.~\ref{fig:Critical couplings for three states}.  We compare the resulting fit with 7 additional points
in the range $5 \leq K < 12$ to illustrate that the fit function also performs reasonably well in this range.  

We examined the light-front wave functions for the lowest state of the odd sector just below and just above the critical couplings 
at each $K$.  We found that they remained dominantly single-boson in character.  By proceeding further beyond the critical 
coupling we find states with more complex character crossing this single-particle dominated state to become the lowest-lying state.  
The details of these transitions in the region beyond critical coupling will be presented in a separate work~\cite{Huang2021}.

\subsection{Lowest state of the even sector}
\label{subsec:lowesteven}

We anticipate that the results for the lowest state of the even sector will likely follow a pattern related to results 
for the lowest state in the odd sector since $\phi^4$ in 1+1 dimensions is known to have no even bound states~\cite{Spencer1974}.  
Thus, we suppose that the mass of the lowest state of the even sector at $K$, which is expected~\cite{hv87} and found to be dominated by 2-boson
configurations, will be close to twice the mass of the lowest state in the odd sector at $K/2$, which is dominated
by the 1-boson configuration.   
We define $f_2(z)$ to be the function describing the 
vanishing mass gap of the lowest state of the even sector and we introduce 
``$R_{2/1}$'' for the ratio (``$R$'') of the mass of the lowest state of the even sector (``$2$'') to the mass of the
lowest state of the odd sector (``$1$''). 
\begin{figure}[ht!]
\centering
\includegraphics[width=15.65cm]{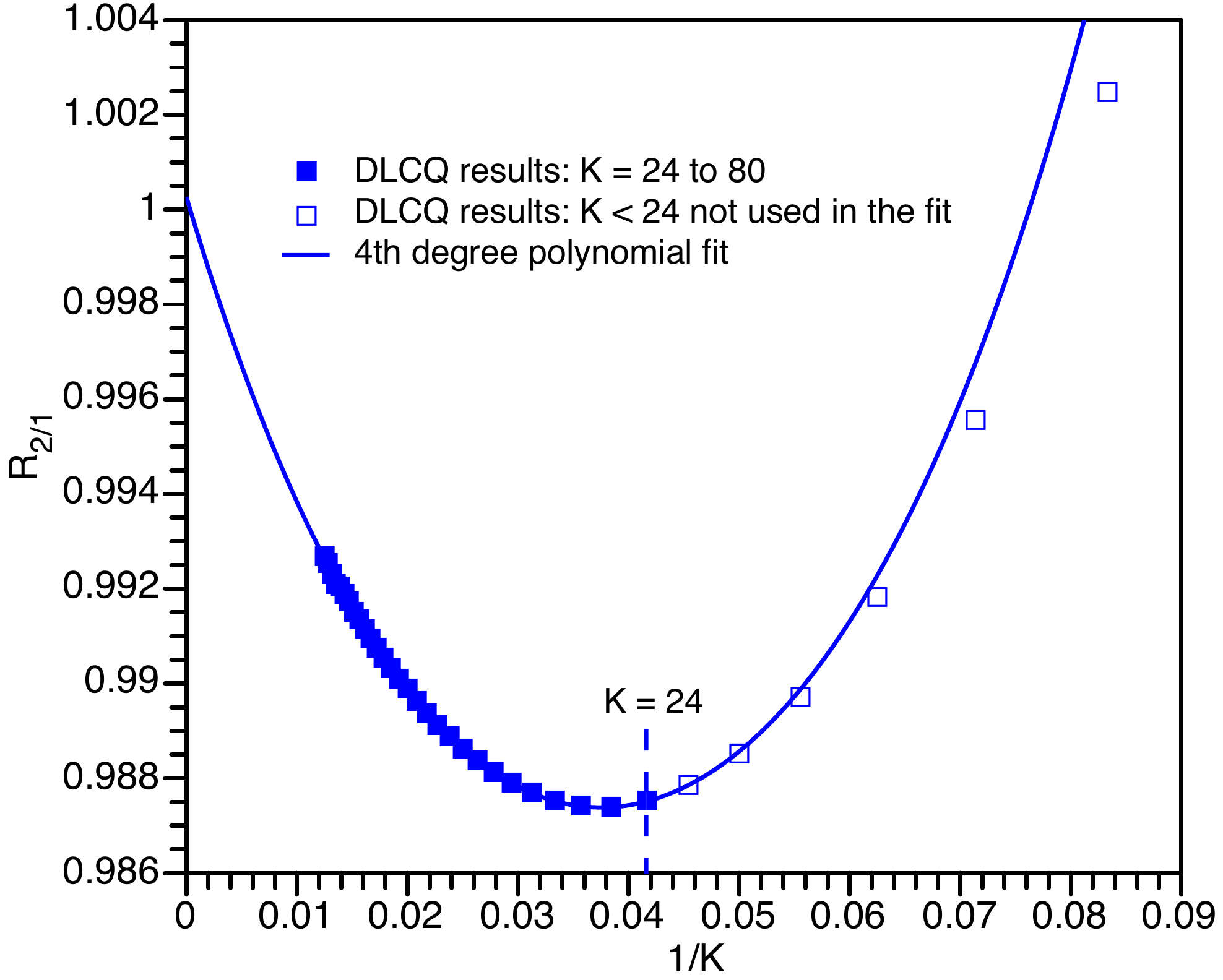}
\caption{Ratio of critical coupling for the vanishing mass gaps of the lowest state at $K$ of the even sector and 
the lowest state of the odd sector at $K/2$ versus $1/K$.  Symbols represent the results from DLCQ and the 
smooth curve is a fourth degree polynomial in $z^2=1/K$ fit to the 29 results represented by the solid symbols.
The coefficients of the fit $B_j$ (see Eq. \ref{Eq:R21aspoly}) are presented in Table \ref{Tab:Expansion Coefficients}. 
The vertical dashed line signifies
the boundary between regions where finite basis artifacts are strong (to the right for lower $K$) yet tending 
to cancel in this ratio as observed by the smooth behavior over all $K$ presented. }
\label{fig:Even to odd critical coupling ratio}
\end{figure} 
We formalize what we anticipate by suggesting that 
\begin{eqnarray}
R_{2/1}(z) = f_2(z)/f_1(\sqrt{2}z)
\label{Eq:R21asratio}
\end{eqnarray}
will produce a smooth function with reduced finite basis artifacts. 
To examine this supposition, we plot the DLCQ results
for the ratio $R_{2/1}$ in Fig.~\ref{fig:Even to odd critical coupling ratio} as a function of $z^2=1/K$. 
We explored different variables for the horizontal axis of this plot and found that $z^2=1/K$ produced results 
that we could then fit with good precision by a 4th-degree polynomial in $1/K$ while other variable choices 
required higher-degree polynomials for fits of comparable quality.
In other words, we define the fit function for the ratio $R_{2/1}$ through
\begin{eqnarray}
R_{2/1}(z)=& \sum_{j} B_j  z^j, 
\label{Eq:R21aspoly}
\end{eqnarray}
where $j$ runs from zero to $8$ in increments of 2.  
For consistency in the treatment of finite $K$ artifacts, the lower limit in $K$ for the even sector DLCQ data 
is chosen so as to correspond with the chosen lower limit $K=12$ used for the DLCQ data adopted for the fit for $f_1$.
Therefore, the DLCQ data for the lowest state of the even sector included in the fit are the 29 points 
ranging over $24 \leq K \leq 80$ with increments of 2 (solid symbols in Fig.~\ref{fig:Even to odd critical coupling ratio}). 
The best fit produces the smooth curve in Fig.~\ref{fig:Even to odd critical coupling ratio} and the residuals for this fit 
are presented below in Fig. \ref{fig:R_Residuals}. 

We tabulate the coefficients $B_j$ of the fit in Eq.~\ref{Eq:R21aspoly} in Table \ref{Tab:Expansion Coefficients}.
Note that the coefficient $B_0 \approx 1.000267$ provides the ratio of our extrapolations for $f_2$ to $f_1$ in the continuum limit.
The difference of $B_0$ from unity fits within our overall uncertainty for the critical coupling in the continuum limit (see below).

Based on the polynomial fits for $R_{2/1}(z)$ and $f_1(z)$, we can obtain $f_2(z)$ using Eq. \ref{Eq:R21asratio}:
\begin{eqnarray}
f_2(z)=R_{2/1}(z)f_1(\sqrt{2}z).
\label{Eq:f_2}
\end{eqnarray}
We present the curve for $f_2(z)$ from this approach in Fig.~\ref{fig:Critical couplings for three states} along
with the DLCQ results for the lowest state of the even sector.  We note that, in order to facilitate comparisons, 
the results in Fig.~\ref{fig:Critical couplings for three states} are presented as functions of $z = 1/\sqrt{K}$.

Since the DLCQ data for the ratio $R_{2/1}$ (including the data excluded from the fit and represented as open symbols in Fig.~\ref{fig:Even to odd critical coupling ratio}) 
appear to be a smooth function of $1/K$, we can conclude that finite basis artifacts appear to be reasonably cancelled when forming this ratio.  
This cancellation facilitates a good fit with a 4th-degree polynomial over the range in $z^2 = 1/K$ that covers $24 \leq K \leq 80$.  Furthermore, the resulting fit 
reasonably describes the DLCQ results below $K=24$ that are not included in the fit.    
For ease of visualization, the $K=24$ boundary is represented by a vertical dashed line in Fig.~\ref{fig:Even to odd critical coupling ratio}. 
We note that this $K=24$ boundary occurs near the point of inflection in $R_{2/1}$ but we do not find
this inflection point to be physically significant.

We test the finite basis artifact cancellations further by comparing the fit displayed in Fig.~\ref{fig:Even to odd critical coupling ratio}
with a fit that includes all the points presented in the Fig.~\ref{fig:Even to odd critical coupling ratio}.  Such an expanded fit goes smoothly through all the points
and is nearly indistinguishable from the curve in Fig.~\ref{fig:Even to odd critical coupling ratio} for the region above $K=24$ (to the left of the vertical dashed line).  
For comparison, the $K \rightarrow \infty$ point changes from $\approx$ 1.0002668 ($B_0$ in Table \ref{Tab:Expansion Coefficients}) to $\approx$ 1.0000843 when all points in Fig.~\ref{fig:Even to odd critical coupling ratio} are included in the fit.

The results shown in Fig.~\ref{fig:Even to odd critical coupling ratio} are predominantly below unity but approach unity in
the continuum limit.  This indicates that, using our connection between the lowest state of the even sector
and the lowest state of the odd sector, the effect of finite $K$ is to produce a bound state in the lowest
state of the even sector at the critical coupling for its vanishing mass gap. Thus, at finite $K$, the lowest
even state, when massless, cannot decay to two of the lowest massless states at $K/2$ of the odd sector.  We checked the 
light-front wave functions for the lowest solution of the even sector near critical coupling at each $K$ 
and confirmed that they are overwhelmingly dominated by the 2-boson configuration.  

It is important to note that our results in Fig.~\ref{fig:Even to odd critical coupling ratio} do indicate that degeneracy 
of the even and odd sector, a condition for spontaneous symmetry breaking, occurs in the continuum limit at the vanishing mass
gaps of both sectors - i.e. the ratio of critical couplings approaches unity as $K \rightarrow \infty$.  In the continuum limit a mix of
lowest  state solutions from the even and odd sector becomes a solution.  In addition, as we see in Sec. \ref{subsec:firstexcited} additional 
degeneracies will occur and will lead to solutions with well-mixed particle content.

\subsection{First excited state of the odd sector}
\label{subsec:firstexcited}

\begin{figure}[ht!]
\centering
\includegraphics[width=16cm]{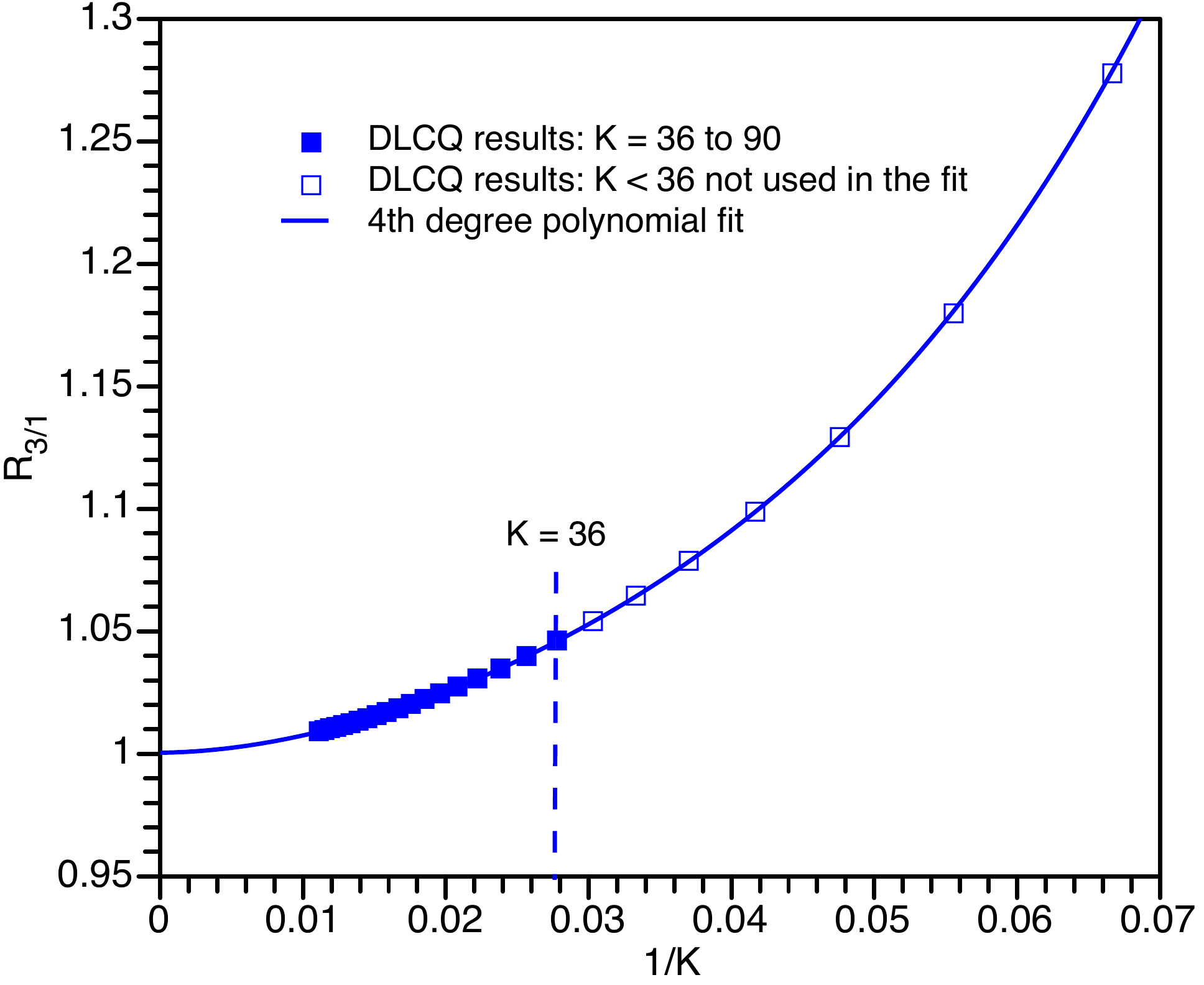}
\caption{Ratio of critical coupling for the vanishing mass gaps of the first excited state at $K$ of the odd sector to 
the lowest state of the odd sector at $K/3$ versus $1/K$.  Symbols represent the results from DLCQ and the 
smooth curve is a fourth degree polynomial in $z^2=1/K$ fit to the 19 results represented by the solid symbols.  
The coefficients of the fit $C_j$ (see Eq. \ref{Eq:R31aspoly}) are presented in Table \ref{Tab:Expansion Coefficients}. 
The vertical dashed line signifies
the boundary between regions where finite basis artifacts are strong (to the right for lower $K$) yet tending 
to cancel in this ratio as observed by the smooth behavior over all $K$ presented.}
\label{fig:First excited to lowest state critical coupling ratio}
\end{figure}

Returning to the odd sector, we consider the critical coupling for the vanishing mass gap of the first excited state
as a function of z which we define as $f_3(z)$.  Here we encountered more severe computational challenges owing to
the higher density of states near the vanishing mass gap for this state.  We sought and found an effective shortcut which
is further detailed below in Sec.~\ref{IMAU}.  In brief, we found that for $K = 75$ the results with a 16-boson truncation
were indistinguishable, within numerical uncertainties, from results without truncation where up to 75 bosons were included.
To conserve computational resources, we therefore retained the 16-boson truncation for DLCQ results at $K \geq 75$.
Examples of the savings in terms of matrix dimension are presented in Table \ref{Table_of_dimensions}.

Since we anticipate and find that this state is dominated by 3-boson configurations,
we extend the logic used above to now consider the ratio of the vanishing mass gap for this state to the vanishing 
mass gap for the lowest state of the odd sector. Hence, we define this ratio as 
\begin{eqnarray}
R_{3/1}(z) = f_3(z)/f_1(\sqrt{3}z).
\label{Eq:R31asratio}
\end{eqnarray}
We plot the DLCQ results for this ratio in Fig.~\ref{fig:First excited to lowest state critical coupling ratio} as
a function of $z^2 =1/K$.  Again, this choice of variable produces a sequence of results that is well-described
by a 4th-degree polynomial
\begin{eqnarray}
R_{3/1}(z)=& \sum_{j} C_j  z^j, 
\label{Eq:R31aspoly}
\end{eqnarray}
where $j$ runs from zero to $8$ in increments of 2 with the resulting smooth curve displayed in 
Fig.~\ref{fig:First excited to lowest state critical coupling ratio}.  Our fit includes 19 DLCQ points over the 
range $36 \leq K \leq 90$ in increments of 3 displayed as solid symbols in Fig.~\ref{fig:First excited to lowest state critical coupling ratio}. 
Note that we attain $K = 90$  for the first excited state of the odd sector owing to the boson number truncation discussed above.
The choice of the lower limit in $K$ for this state follows a similar line of reasoning for consistent treatment of finite $K$ artifacts 
that we invoked for the lowest  state of the even sector in Sec.~\ref{subsec:lowesteven}.

We tabulate the coefficients $C_j$ of the fit in Eq.~\ref{Eq:R31aspoly} in Table \ref{Tab:Expansion Coefficients}.
Note that the coefficient $C_0 \approx 1.000463$ provides the ratio of our extrapolations for $f_3$ to $f_1$ in the continuum limit.
The difference of $C_0$ from unity also fits within our overall uncertainty for the critical coupling in the continuum limit (see Sec. \ref{sec:R&UQ}).

We now offer an observation about finite basis artifacts analogous to our previous observation concerning $R_{2/1}$ 
in Fig.~\ref{fig:Even to odd critical coupling ratio}.
That is, it is interesting to note that finite basis artifacts appear again to
be well-suppressed in the ratio $R_{3/1}$ since the 4th-degree polynomial represents DLCQ results accurately over a range 
in $z^2 = 1/K$ that extends well below the $K = 36$ boundary  in Fig.~\ref{fig:First excited to lowest state critical coupling ratio}
that separates the results employed in the fit (solid symbols) from those excluded (open symbols).

We are now in a position to obtain $f_3(z)$,  based on our pair of polynomial fits 
\begin{eqnarray}
f_3(z)=R_{3/1}(z)f_1(\sqrt{3}z).
\label{Eq:f_3}
\end{eqnarray}
We present the curve for $f_3(z)$ from this approach in Fig.~\ref{fig:Critical couplings for three states} along
with the DLCQ results for the first excited state of the odd sector.  

Following our exploration of the finite basis artifact cancellations for $R_{2/1}$ we compare the fit displayed in Fig.~\ref{fig:First excited to lowest state critical coupling ratio}
for $R_{3/1}$ with a fit that includes all the points presented in the Fig.~\ref{fig:First excited to lowest state critical coupling ratio}.  
Such an expanded fit goes smoothly through all the points and is nearly indistinguishable from the curve in Fig.~\ref{fig:First excited to lowest state critical coupling ratio} for the entire range of $1/K$ displayed.
For comparison, the $K \rightarrow \infty$ point changes from $\approx$ 1.0004631 ($B_0$ in Table \ref{Tab:Expansion Coefficients}) to $\approx$ 1.0015136 when all points in Fig.~\ref{fig:First excited to lowest state critical coupling ratio} are included in the fit.

Unlike the ratio $R_{2/1}$, the ratio $R_{3/1}$ remains everywhere above unity at finite $K$ while approaching
unity smoothly and monotonically as $K \rightarrow \infty$.  This indicates that the first excited state of the odd
sector at finite $K$, when approaching a vanishing mass, is able to decay to three separated interacting bosons each carrying K/3 units of the 
longitudinal momentum of the parent state.  Unity in the asymptote of $R_{3/1}$ signals degeneracy of the 3-boson
dominated state with the 1-boson dominated state at a vanishing mass gap for both in the continuum limit.  

Up to this point, as summarized in Fig.~\ref{fig:Critical couplings for three states}, the DLCQ results indicate the
anticipated degeneracy of the even and odd sectors and the lowest and first excited states of the odd sector
in the continuum limit $K \rightarrow \infty$ all with a vanishing mass gap.  These simultaneous transitions
are found to occur with a critical coupling of $22.64\pm 0.17$.
%\textcolor{red}{ $22.64\pm 0.17$ (preliminary)}. 
Our quoted uncertainty
is an estimated systematic uncertainty which is expected to dominate our overall uncertainty as discussed in Sec.~\ref{SPFF}.
For comparison, the corresponding critical coupling result 
of Ref. \cite{bch} (\cite{agkkw}) is 26.39 $\pm 0.63$ (23.12 $\pm 0.38$).
Hence, taking the respective quoted uncertainties into account, we are consistent with the results of 
Ref.~\cite{agkkw} as depicted in the inset to Fig.~\ref{fig:Critical couplings for three states}.

\section{Residuals and Uncertainty Quantification}
\label{sec:R&UQ}

There are a number of sources of uncertainties in our implementation of  DLCQ.  An abbreviated list includes:

\begin{enumerate}
\item{Hamiltonian mass eigenvalue calculations}
\item{Interpolation for the vanishing mass gap}
\item{Selection of functions and variables for fitting critical couplings as a function of $K$}.
\item{Role of residual finite $K$ artifacts}
\end{enumerate}

Readers less interested in the details of the uncertainty analysis may wish to either scan subsection \ref{SPFF} 
which details the largest source of our uncertainty or skip this section entirely.

\subsection{Uncertainties in eigenvalues}
\label{UIE}

We have mentioned above that we have developed, tested and employed three independent codes.
Two of these codes,  ``lfphi4MH-00''~\cite{Mengyao_code} and ``lfphi4SJ-00''~\cite{Shreeram_code}, 
are special purpose codes that have produced our mass eigenvalues in double-precision (64-bit accuracy) for $K \leq 42$.
They have been benchmarked with each other to verify their eigenvalues agree within double-precision. 

A third code (MFD)~\cite{mfd,chmv,chmpv,chv} is a Fortran77/Fortran90 code which is MPI parallelized.  This code, which has been enabled to solve
boson systems, employs the Lanczos algorithm to obtain the lowest mass-squared eigenvalues in single-precision (32-bit accuracy). 
We note that MFD employs double-precision for key components of the Lanczos 
iteration process in order to maintain overall single-precision accuracy.  Benchmarking the three codes for the region $K \leq 42$
where all three run successfully, confirms MFD produces results with single-precision accuracy.  MFD alone is employed to obtain
our results in the range $K>42$.

In the region where the mass gap vanishes, the mass-squared eigenvalues are typically less than unity.  In this region we find that MFD's
lowest eigenvalue has an uncertainty of about unity in the fifth decimal place.
Support for the assertion of MFD's accuracy 
at larger values of $K$, which most strongly influences our extrapolations,  will be seen below in our presentation of detailed results at $K=75$ 
in Sec. \ref{IMAU}.  

%\textcolor{red}{Additional discussion from Mamoon's eigenvalue uncertainty analysis}

\subsection{Interpolation method and uncertainties}
\label{IMAU}

\begin{figure}[ht!]
\centering
\includegraphics[width=16cm]{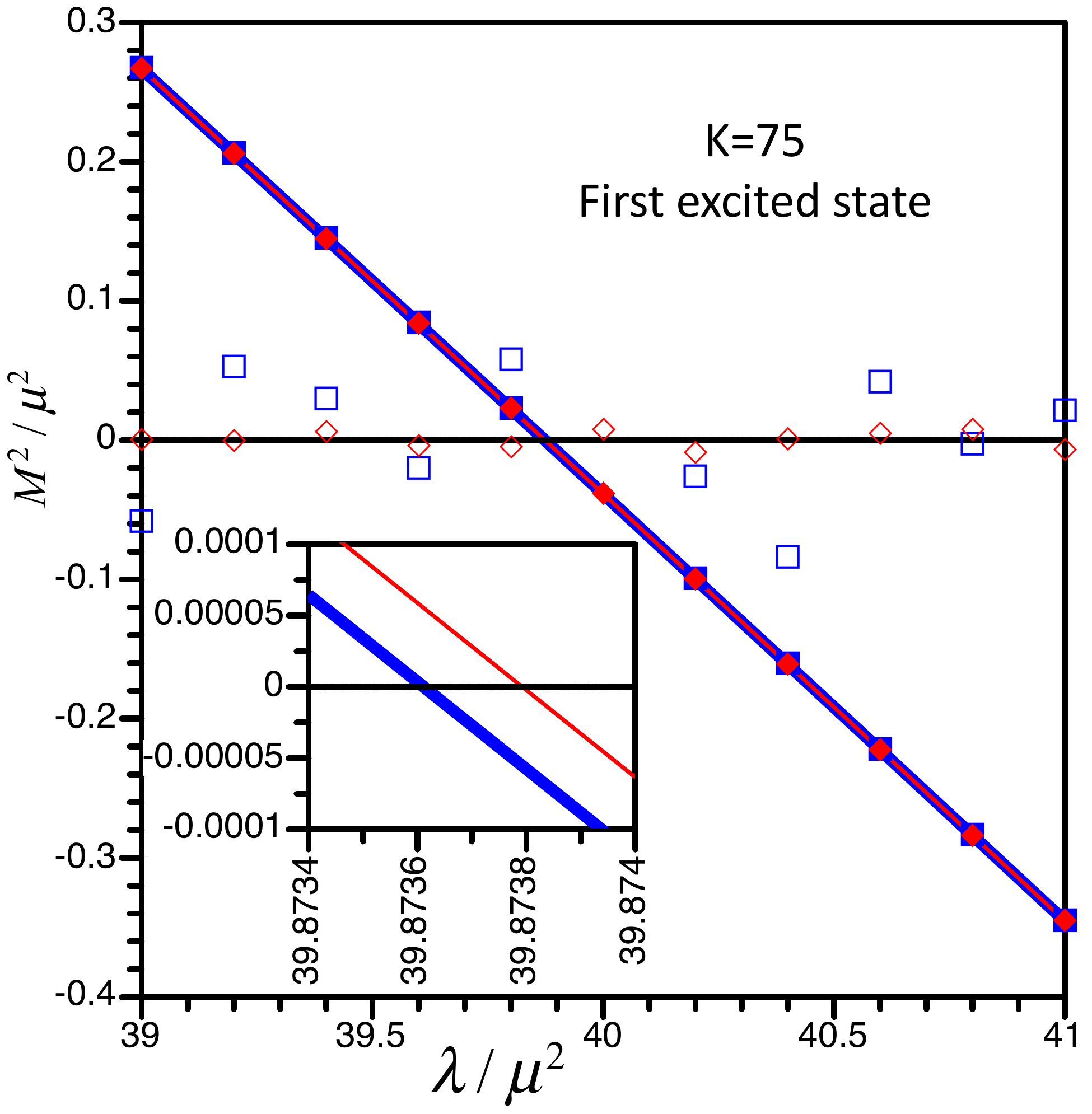}
\caption{Calculated $M^2/\mu^2$ eigenvalues (solid symbols) for the first excited state in the odd sector for couplings that span the vanishing mass gap at $K=75$.
The solid blue squares correspond to the full calculation while the solid red diamonds correspond to results in a reduced space that omits configurations
with more than 15 bosons.
Results are presented for increments of 0.2 in the coupling except for one point in the full calculation at 40.0, just above the critical coupling, 
where 7500 Lanczos iterations were insufficient for producing a converged result.
The thick blue line (thin red line) corresponds to a second degree polynomial fit used to interpolate for the vanishing mass gap.
The open symbols present 1000 times the difference between the corresponding solid symbol and fit.  
The inset provides an expanded view of the two quadratic fits as they cross the vanishing mass gap. As expected, the coupling must be larger
for the truncated calculation relative to the full calculation to produce a vanishing mass gap.}  
\label{fig:X_Interpolation_uncerts}
\end{figure}

In order to obtain the critical coupling for a vanishing mass gap at a specific value of $K$, we perform a sequence of
calculations at values of the coupling adjacent to the vanishing mass gap. 
We perform these calculations over a narrow range of coupling separately for each of our three states 
since different values of the critical coupling will emerge for each state at each value of $K$.  
This is evident from taking any vertical line through the final results presented in Fig.~\ref{fig:Critical couplings for three states} 
and projecting the intercepts to the vertical axis.

For $K \leq 42$ we use ``lfphi4MH-00'' to search for the value of the critical coupling to eight significant figures producing a vanishing mass gap.
For $K > 42$ we use MFD to produce results on a mesh of values of the coupling that span the vanishing mass gap and obtain the 
critical coupling by interpolation as described below.

We choose the $K=75$ example to present our procedure in more detail.  In addition, we choose the most challenging state to converge 
(requiring the largest number of Lanczos iterations), the first excited state of the odd sector for this detailed study.  In fact, we already
experienced Lanczos convergence difficulties when attempting full calculations for this first excited state in the odd sector at a 
value of the coupling close to the critical coupling (see below). This led us to consider implementing a boson number truncation at 16 bosons 
for DLCQ basis states in the range $K>75$.

To be specific, the DLCQ eigenvalues for the full calculation of the first excited state in the odd sector at $K=75$ in 
Fig.~\ref{fig:Critical couplings for three states} were obtained with 7500 Lanczos iterations.  
Beyond 7500 Lanczos iterations we experienced numerical difficulties for this first excited state.  Note that even more iterations 
would be needed at higher $K$ to achieve convergence.  By contrast, 1600 Lanczos iterations are sufficient to converge 
the lowest state with a full basis in the odd sector at couplings near the vanishing mass gap even at $K=88$, the highest $K$ we achieved with a full basis.

We present the DLCQ $K=75$ results for the first excited state in the odd sector as solid symbols in Fig.~\ref{fig:X_Interpolation_uncerts} 
on an evenly spaced mesh that spans the vanishing mass gap.  We chose 11 points about evenly distributed above and below what we estimated as the critical coupling
with increments of 0.20.  We present two cases in the figure: the full calculation (solid blue squares) that keeps all boson configurations up to and including the 75-boson
configuration and a calculation where only states with 16 or fewer bosons were retained (solid red diamonds). The iterations failed to converge the eigenvalue at  $\lambda$ = 40.0
in the full calculation.  Otherwise, the eigenvalues from the two calculations are indistinguishable by eye on this scale.

We then fit the results of each calculation with a second-degree polynomial which we found suitable for extracting the critical coupling by interpolation.  
Fig.~\ref{fig:X_Interpolation_uncerts} then presents the residuals between each of the fits and the corresponding calculated points as open 
symbols with the corresponding shape and color.  In order to render the displacement of the residuals from zero more visible, 
we multiply the residuals by a factor of 1000.  

The residuals appear to be reasonably distributed about zero which suggests the adequacy of our choice of a second degree polynomial as the function
for fits and interpolations.  Furthermore, as may be anticipated from the reduction in the Hamiltonian matrix size by more than a factor of 2 
from 4,059,416 for the full calculation to 1,824,575 for the matrix truncated to the limit of 16-boson configurations, the residuals are noticeably reduced
with that truncation.  The resulting rms deviation between calculation and fit is $4.56 \rm{x} 10^{-5}$  ($5.61\rm{x}10^{-6}$) for the full calculation with 10 points 
(16-boson truncation with 11 points). The resulting uncertainty in the critical $\lambda$ at $K = 75$ is obtained by incorporating information from the slope of the quadratic fit function.  
With this we arrive at an uncertainty in the critical $\lambda$ at $K = 75$ of $1.57  \rm{x} 10^{-4}$ ($1.83  \rm{x} 10^{-5}$) for the full (16-boson truncation) calculation.

In order to examine the difference in the interpolated critical coupling, we provide an inset to Fig.~\ref{fig:X_Interpolation_uncerts} with an expanded
scale.   The inset shows the two fit functions provide vanishing mass gaps that differ by about 0.00002 in their critical couplings. As expected, 
a larger coupling is needed with the truncated matrix to produce a vanishing mass gap.  This uncertainty is small compared to other uncertainties that we will assess below.

Referring back to the discussion in Sec.~\ref{UIE}, we can also use these deviations presented in Fig.~\ref{fig:X_Interpolation_uncerts} to infer an eigenvalue uncertainty.
That is, taking the deviations between calculation and fits as a gauge, we estimate that eigenvalues at the extrema in this figure, on average, are accurate to about 4 (5) decimal places 
for the full (truncated) calculation.  

We conclude this discussion by noting that we explored changing the range of the couplings spanning the vanishing mass gap.  We also explored increasing the number of points calculated for the fitting and interpolation process. This led us to estimate that our uncertainty in the deduced critical coupling at each fixed $K$ and for each of the three 
states studied could conservatively be assigned a value of unity (two units) in the sixth (fourth) decimal place for $K \leq 42$ ($K > 42$).  
We note that, for  $K > 42$ this error estimate is 
the same as the systematic shift in the critical coupling shown to arise from truncation to 16-boson configurations 
in Fig.~\ref{fig:X_Interpolation_uncerts}.  Hence, at this stage of the uncertainty analysis, our dominant uncertainty 
in deduced critical couplings is less than five units in the fourth decimal place.

\subsection{Selection of polynomials for fitting}
\label{SPFF}

\begin{table}[ht]
\begin{centering}
\begin{tabular}{|c|c|c|c|c|}
\cline{2-5} 
\multicolumn{1}{c|}{} & \multicolumn{4}{c|}{Extrapolations using Polynomials of Varying Degree }\tabularnewline
\hline 
Max $K$ & $3^{\rm rd}$ &  $4^{\rm th}$  & $5^{\rm th}$  & $6^{\rm th}$  \tabularnewline
\hline 
\hline 
$\mbox{64}$ & 22.43333	& 22.86767	& 22.59842	& 22.81205  \tabularnewline
\hline 
$\mbox{68}$ & 22.45172	& 22.85482	& 22.61448	& 22.81022  \tabularnewline
\hline 
$\mbox{72}$ & 22.46768	& 22.84330	& 22.62603	& 22.80189  \tabularnewline
\hline 
$\mbox{76}$ & 22.48159	& 22.83222       & 22.63073	& 22.77323  \tabularnewline
\hline 
$\mbox{80}$ & 22.49397	& 22.82240	& 22.63476	& 22.75340  \tabularnewline
\hline 
$\mbox{84}$ & 22.50107	& 22.81590	& 22.63640	& 22.73715 \tabularnewline
\hline 
$\mbox{88}$ & 22.50877	& 22.80870	& 22.63806	& 22.72247  \tabularnewline
\hline 
\end{tabular}
\par\end{centering}

\caption{Extrapolations of $\lambda_c$ for the lowest state of the odd sector as a function of the degree of the polynomial adopted and as a function
of the Max $K$ included in the data set. The overall range is $K$ = $12 - 88$ for a maximum of 52 DLCQ points. 
The extrapolations of the $5^{\rm th}$ degree polynomial exhibit the least sensitivity to the Max $K$ over
the range of Max $K$ values displayed in the table.}
\label{Tab:Sensitivity_to_Degree}
\end{table}

We first consider the DLCQ results for $f_1(z)$ shown in Fig.~\ref{fig:Critical couplings for three states} and represented by Eq.~\ref{Eq:f_1}.
The choice of variable $z=1/\sqrt{K}$ emerged after attempts with $z$ to a variable power revealed that a robust low-degree polynomial fit could be obtained with 
a polynomial in $z$ itself.  The meaning of ``robust'' will become more apparent shortly.  

Among several methods of partitioning our DLCQ results for fitting and extrapolating, we present in Table~\ref{Tab:Sensitivity_to_Degree} 
an example of extrapolations to the continuum limit ($K \rightarrow \infty$ or $z  \rightarrow 0$) using a sequence of polynomial degrees.  
In particular, we extrapolate the critical coupling for the lowest eigenstate of the odd sector to the continuum limit using a range of polynomial degrees
(column labels) and a set of cutoffs in $K$ retained for the fit (``Max $K$'').  For a third (fourth) degree polynomial, the extrapolation steadily increases 
(decreases) with increasing Max $K$.  The fifth degree polynomial appears to approximately average the third and fourth
degree result at each value of Max $K$ including the highest Max $K=88$.  The sixth degree fit starts out on a plateau at lower values of Max $K$ and then
falls towards the result of the fifth degree polynomial at Max $K=88$.  It is interesting to note that the average of the four extrapolations
at Max $K=88$ in Table~ \ref{Tab:Sensitivity_to_Degree} is 22.66950 which is within 0.03144 of the result of the fifth degree polynomial.
Owing to the stability of its extrapolation as a function of Max $K$, and to the desire to avoid overfitting our limited data set, 
we adopt the fifth degree polynomial for our overall choice and use it for further sensitivity studies.

Having fixed our attention on the initial indication of reasonably stable extrapolations (i.e. stable with increasing Max $K$) from the fifth degree polynomial, 
we continue to test for a robust character by studying the
extrapolation for $f_1(z)$ as a function of various upper and lower cutoffs in the DLCQ results retained in the fit.  We present results for $f_1(z)$ extrapolations 
employing various selections of DLCQ data sets in Table~\ref{Tab:Extraps_from_trimfits}.  For the extrapolations corresponding to listed values of Max $K$, 
the minimum value of $K=12$.  For the extrapolations corresponding to listed values of Min $K$, the maximum value of $K=88$.

\begin{table}[ht]
\begin{centering}
\begin{tabular}{|c|c|c|c|}
\cline{2-4} 
\multicolumn{1}{c|}{} & \multicolumn{3}{c|}{Extrapolations from $5^{\rm th}$ Degree Polynomial Fits }\tabularnewline
%\multicolumn{1}{c|}{} &
%\multicolumn{4}{Extrapolations from $5^{\rm th}$ Degree Polynomial Fits }\tabularnewline
\hline 
Max $K$ & $\lambda_c$  & Min $K$  & $\lambda_c$  \tabularnewline
\hline 
\hline 
$\mbox{48}$ &  22.49531  & $\mbox{52}$ & 25.48713  \tabularnewline
\hline 
$\mbox{52}$ & 22.53590  & $\mbox{48}$ & 23.64028   \tabularnewline
\hline 
$\mbox{56}$ & 22.56649 & $\mbox{44}$ & 22.28556   \tabularnewline
\hline 
$\mbox{60}$ & 22.58336 & $\mbox{40}$ & 22.51619   \tabularnewline
\hline 
\hline
$\mbox{64}$ & 22.59842 &  $\mbox{36}$ & 22.60208  \tabularnewline
\hline 
$\mbox{68}$ & 22.61448 & $\mbox{32}$ & 22.54641   \tabularnewline
\hline 
$\mbox{72}$ & 22.62603 & $\mbox{28}$ & 22.56581  \tabularnewline
\hline 
$\mbox{76}$ & 22.63073 & $\mbox{24}$ & 22.61545  \tabularnewline
\hline 
$\mbox{80}$ & 22.63476 & $\mbox{20}$ & 22.65623  \tabularnewline
\hline 
$\mbox{84}$ & 22.63640 & $\mbox{16}$ & 22.66967  \tabularnewline
\hline 
$\mbox{88}$ & 22.63806 & $\mbox{12}$ & 22.63806  \tabularnewline
\hline 
\end{tabular}
\par\end{centering}

\caption{Extrapolations of $\lambda_c$ to the continuum limit ($K \rightarrow \infty$) from $5^{\rm th}$ degree polynomial fits ($f_1$ of Eq. \ref{Eq:f_1}) using ranges 
of DLCQ results specified by the upper value (Max $K$) and the lower value (Min $K$) of the range.  The last row presents the result when all 52 DLCQ results
over the range $12 \leq K \leq 88$ are employed.}
\label{Tab:Extraps_from_trimfits}
\end{table}

By sliding the upper cutoff down from Max $K=88$ we find comparable extrapolations to those shown already in Table~ \ref{Tab:Sensitivity_to_Degree} 
down as far as Max $K=56$ (38 points retained in the fit).  Less sensitivity is obtained by sliding Min $K$ up from 12 (all 52 points included in the fit) toward 
Min $K=36$ (28 points retained in the fit).  Going to the higher values of Min $K$ that are listed produces significantly larger deviations in the extrapolations.

We estimate that the extrapolations using Max $K$ $\geq 64$ and Min $K$ $\leq 36$, i.e. those below the double line in Table~\ref{Tab:Extraps_from_trimfits},  
indicate that the fifth degree polynomial is a robust choice. The mean value of these 13 extrapolations (using the entry on the bottom row just once) is 22.618 
with a maximum deviation $\approx$ 0.072. We will observe that this is close to our final result. 

Based on these studies  we see that the systematic uncertainties from the choice of the degree of the polynomial (Table~\ref{Tab:Sensitivity_to_Degree}) and to
the choice of the range of data included in the $5^{\rm th}$ degree polynomial fit (Table~\ref{Tab:Extraps_from_trimfits}) dominate our overall uncertainties 
compared to those estimated above.  
To be conservative, we quote our critical coupling and its associated uncertainty so as to encompass the results in the last row of Table~\ref{Tab:Sensitivity_to_Degree} 
as well as all those in Table~\ref{Tab:Extraps_from_trimfits} below the double line.  This provides our final result for the critical coupling:  $22.64\pm 0.17$.
% \textcolor{red}{  $22.64\pm 0.17$ (preliminary)}.  
 By electing the single overall uncertainty to encompass these selected results in Tables~\ref{Tab:Sensitivity_to_Degree} and \ref{Tab:Extraps_from_trimfits}, 
 we estimate that our smaller uncertainties, discussed above, are well covered.

\subsection{Residual finite $K$ artifacts}

\begin{figure}[ht!]
\centering
\includegraphics[width=16cm]{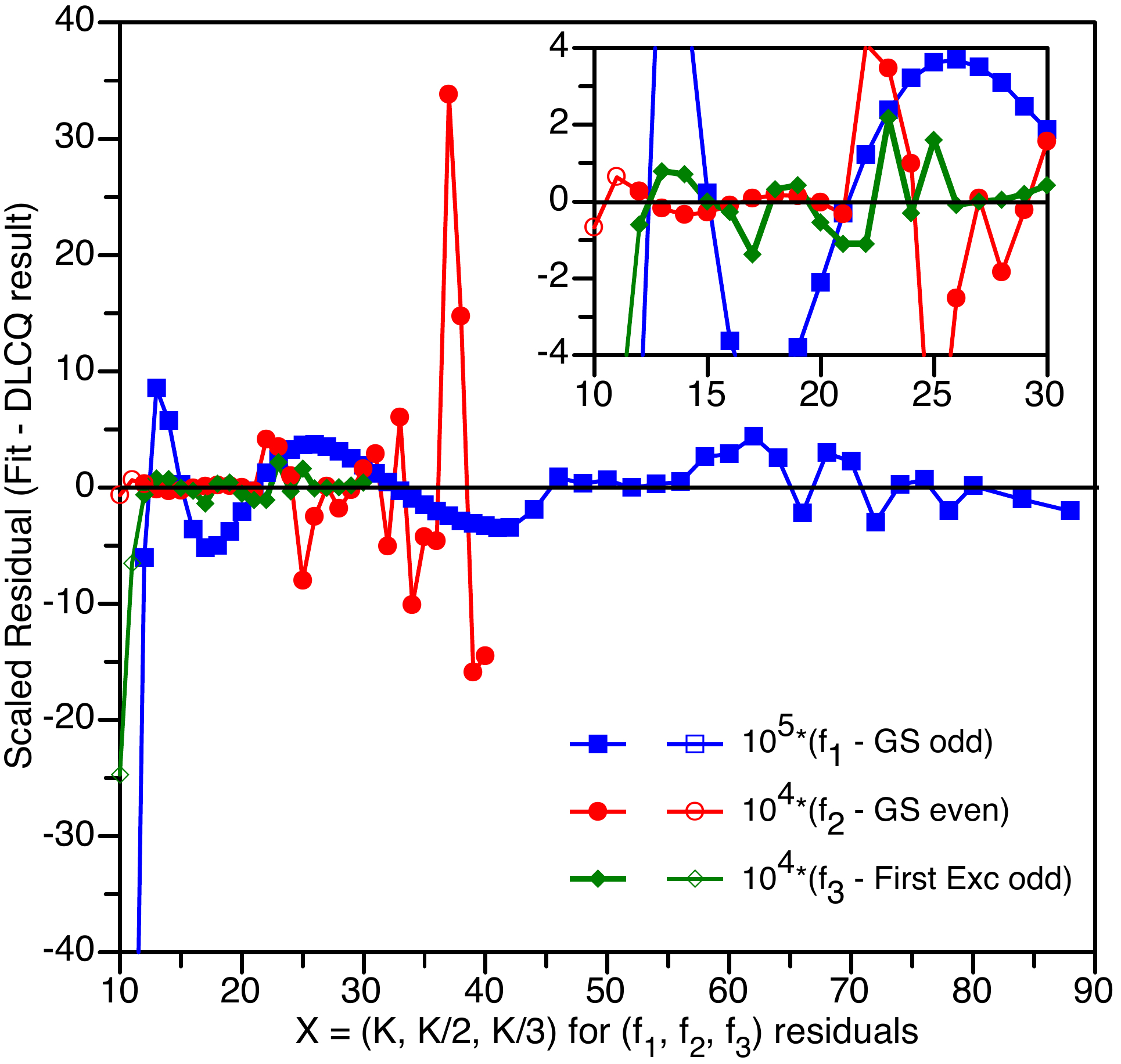}
\caption{Residuals, plotted as functions of $X=(K, K/2, K/3)$ for $(f_1, f_2, f_3)$ respectively,
 between functions based on polynomial fits and DLCQ mass-squared eigenvalues. 
That is, these are the residuals of the results presented in Fig. \ref{fig:Critical couplings for three states}.
The residuals are scaled, as indicated by the scaling factors in the legend, to generate visible results on a single vertical scale. 
Filled (open) symbols correspond to DLCQ results employed (not employed) in the fits. Connecting straight line segments
are included to help guide the eye.
The horizontal axis $X$ is defined for each function according to the points from $f_1$ employed in its evaluation (see text for details).
The inset exhibits details of the residuals in the lower-X region on an expanded scale.
}
\label{fig:f_Residuals}
\end{figure}

\begin{figure}[ht!]
\centering
\includegraphics[width=16cm]{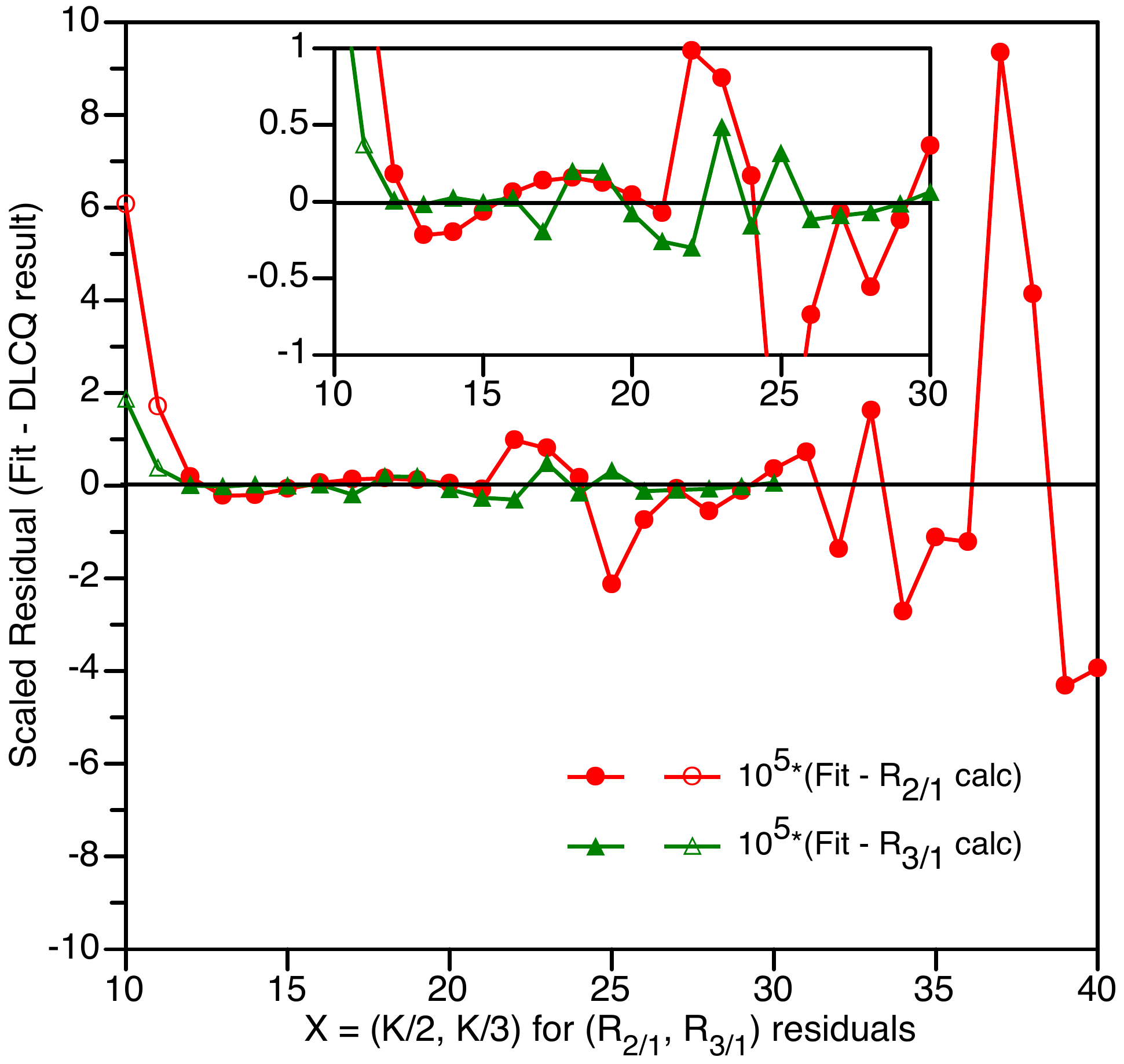}
\caption{Residuals, plotted as functions of $(K/2, K/3)$ for $(R_{2/1}, R_{3/1})$ respectively,
 between polynomial fits and calculated ratios of mass-squared eigenvalues.  
The residuals are scaled, as indicated by the scaling factors in the legend. 
The ratio $R_{2/1}$ ($R_{3/1}$) is defined in the text 
by Eq.~\ref{Eq:R21asratio} (Eq.~\ref{Eq:R31asratio}) and displayed in Fig.~\ref{fig:Even to odd critical coupling ratio} (Fig.~\ref{fig:First excited to lowest state critical coupling ratio}).
Filled (open) symbols correspond to DLCQ results employed (not employed) in the fits. Connecting straight line segments
are included to help guide the eye. 
The horizontal axis $X$ is defined for each function according to the points from $f_1$ employed in its evaluation (see text for details).
}
\label{fig:R_Residuals}
\end{figure}

Having obtained the functions representing the DLCQ results, we turn our attention to discussing the residuals, the differences between fit functions
and DLCQ calculations and ratios of DLCQ calculations.
We present the residuals for the functions $f_1, f_2$ and $f_3$ in Fig.~\ref{fig:f_Residuals} scaled by the indicated large factors 
for viewing on the same overall scale.
Here we observe the appearance of the finite basis artifacts embodied in the function $f_1$ as a smooth function for $K \leq 42$.  These artifacts
also play a role in the functions $f_2$ and $f_3$ as represented by Eqs. \ref{Eq:f_2} and \ref{Eq:f_3}.

In light of these relationships of $f_2$ and $f_3$ to $f_1$, our DLCQ results for $f_1$ extend further towards the continuum limit than our
results for the other two functions.  Hence, we focus on the residuals for $f_1$ with the understanding that their narratives, with appropriate changes 
in their arguments, relate strongly to the narrative for $f_1$.  

Viewing the residuals for $f_1$ in Fig.~\ref{fig:f_Residuals}  from low to high $K$, one clearly observes the smooth function for $K \leq 42$ transitions 
to a noisy distribution for $K > 42$. This is as expected since this transition corresponds to the changeover from the region of double-precision results 
to the region of single-precision results with a resulting increase in numerical uncertainties.  
We observe that the largest deviation from zero of an $f_1$ residual is about $8.5 \times 10^{-5}$ on a scale where $\lambda_c$ 
is on the order of $30$ indicating an inferred fractional error of about $ 3 \times 10^{-6}$.  Furthermore, most of the inferred fractional errors 
for $f_1$ are a factor of 2 to 3 smaller than this.  These observations are reasonably consistent with our assertion of single-precision accuracy
in the region $K > 42$.

Since DLCQ results for $f_1$ with a rescaled argument are employed in the determination of ratios of DLCQ results used at the next stage of the 
fitting process, we take this opportunity to present the residuals for $f_2$ and $f_3$  also in Fig.~\ref{fig:f_Residuals}.  Here, we employ a variable
x-axis scale that corresponds to the argument of $f_1$ employed for determining ratios that are then fit for purposes of extrapolations.
Considering the rescaled argument, we expect and find the transition from smooth to noisy finite $K$ artifacts for $f_2$ ($f_3$) occurs at 
$K/2 = 21$ ($K/3$=14).
For $f_3$, however, the noise appears much smaller than the corresponding noise for $f_2$. This is an indication that the
numerical noise for the lowest state and the first excited state in the odd sector are correlated resulting in reduced
$f_3$ residuals extending to the limits of our current calculations.

To be complete, we also present residuals for $R_{2/1}$ and $R_{3/1}$ in Fig.~\ref{fig:R_Residuals} scaled
by the large factors indicated in the legend.  We observe that the patterns of the larger residuals 
for $R_{2/1}$ and $R_{3/1}$ follow the larger residuals for $f_2$ and $f_3$ respectively in Fig.~\ref{fig:f_Residuals}.  
We can trace this approximate repeating pattern back to the results shown for $f_1$ in Fig.~\ref{fig:f_Residuals}.  
In particular, we observe that the residuals for $f_1$ are about an order of magnitude smaller for $K \geq 42$ 
than those for $f_2$ and $f_3$ in their corresponding domains of $K \geq 24$ and $K \geq 36$ respectively
(note the scale factors in the legends).
Hence, the residuals for $R_{2/1}$ and $R_{3/1}$ are dominated by the numerical deviations of calculated points 
for $f_2$ and $f_3$ from their polynomial fits.

\section{Summary and Outlook}

The main result of the present work is that DLCQ is capable of producing 
the critical coupling for the vanishing mass gap in two-dimensional $\phi^4$
theory, with accuracy competitive with other methods. 
This is very important for several reasons. The earlier
work on this problem~\cite{hv87} was done in an era when both DLCQ and
supercomputing were in their infancy. There were many lingering doubts about 
the importance of zero longitudinal momentum modes and the reliability of
DLCQ which 
is conceptually simple and 
employs a uniform grid in momentum space. 
When the exactly zero longitudinal momentum mode is dropped in this theory,
a question is raised about the potential role it could have played
near the critical coupling 
which is associated with the onset of boson condensation. 
Our results show that        
working with Periodic Boundary Condition (PBC) and ignoring the 
zero mode appears to be a valuable and reliable approach and produces results consistent 
with an alternative light-cone approach that 
uses basis functions that span the entire region of momentum fraction 
from 0 to 1 in the continuum~\cite{agkkw}.

Nevertheless, a constrained longitudinal zero mode does exist when 
PBCs are used. Hence, in spite of the excellent agreement with
another light front method which does not employ DLCQ, one may still 
ask how its inclusion will affect the present result for the critical coupling.
As we already mentioned in Sec.  II, incorporating the zero mode 
in DLCQ is a nontrivial problem and to the best of our knowledge,
this problem has not yet been solved in a satisfactory manner. A proposal
on how to address this issue is presented, for example, in Ref. 
\cite{Chabysheva:2009rs}.  Further in-depth investigations are needed 
to settle this issue quantitatively.

Since the critical coupling in this theory is associated with a second order
phase transition from the symmetric phase to the spontaneously 
broken  symmetry  phase, 
one crucial issue is whether DLCQ can detect the clear signal of 
spontaneous symmetry breaking (SSB). In the symmetric phase 
the eigenstates of the Hamiltonian share the symmetry of the Hamiltonian
which  is invariant under $\phi \rightarrow -\phi$. But in the symmetry
broken phase, while the Hamiltonian retains the symmetry, one can find 
eigenstates that do not share this symmetry.
As we show here, DLCQ predicts, to within numerical precision, the degeneracy of the lowest 
state of the odd sector and the lowest state of the even sector in the 
continuum limit at the vanishing mass gap - a key feature of spontaneous 
symmetry breaking. By the linear superposition of these two states,  
it now becomes possible to  construct an eigenstate of the Hamiltonian
which does not share the symmetry of the Hamiltonian - a clear
signature of SSB.

Spectrum degeneracy between odd and even sectors is a feature of the entire 
SSB phase and hence is manifested also for 
couplings above the critical coupling. However, the critical point itself has
other characteristic features. A continuous spectrum is a key feature of the
critical point. Thus the spacing between eigenvalues should also 
vanish~\cite{agkkw}. In the critical region governed by strong interactions, 
it becomes numerically more challenging to access the masses of the
excited states. But, using DLCQ, we are able to establish the vanishing mass
gap also between the lowest two excitations in the odd sector in the continuum
limit, a result previously achieved by Ref.~\cite{agkkw}.  We also observe that, 
away from the critical point, this degeneracy is lifted.     
  
Our final result for the critical coupling is  $22.64\pm 0.17$
%\textcolor{red}{ $22.64\pm 0.17$ (preliminary)} 
where the uncertainty arises overwhelmingly from our systematic uncertainty.
For comparison, the critical coupling result 
of Ref. \cite{bch} (\cite{agkkw}) is 26.39 $\pm 0.63$ (23.12 $\pm 0.38$).

We note that the critical coupling of Ref.~\cite{bch}
employing a polynomial basis, was obtained after a truncation in the  maximum of  bosons 
allowed
in the configurations retained (7 for the odd and 8 for the even
sectors respectively).  In addition they also implemented  
independent tuning of resolutions in each Fock sector. 
We performed 
a set of DLCQ calculations with an 8-boson truncation for $f_1$ and 
followed the same fitting and extrapolation
procedures presented above.  With our 8-boson truncation we obtain a critical 
coupling of $23.85\pm 0.17$.
Thus, it appears that the difference in  results for the critical coupling
between Ref.~\cite{bch} 
and  DLCQ is not due to an 8-boson truncation in DLCQ alone.

There are many open questions which require further investigation.  We
discuss  a few salient examples.

The results presented in this work are made possible with advances both in
code development (software) and access to state of the art supercomputers
(hardware).  By additional investments in code development and by securing 
additional 
computational resources, can we further refine our prediction for the
critical coupling?

Can one demonstrate the vanishing of the wave function renormalization
constant at the critical point~\cite{dhms,hm} in the continuum limit?
Can one compute the critical exponents? They have the virtue that they are
universal, independent of the regularization and the renormalization schemes. 
On the other hand except for the case of the mass gap, they are notoriously 
difficult to calculate. 

How would results change if we adopt Anti-Periodic Boundary Conditions (APBC) 
where both the conceptual and technical problems associated  with 
the constrained  zero mode operator are absent? Unlike the SSB phase, in
the symmetric phase  we
expect the spectrum to be independent of the boundary conditions as the
continuum limit is approached.

Can we reveal the detailed nature of the phase transition by proceeding 
to stronger coupling with PBC?
From our finite but large $K$ results we have observed that in the symmetric phase there is no level mixing in
the low-lying spectrum. That is, invariant masses of the lowest two states
in the odd sector and of the lowest state in the even sector go to zero at 
the critical coupling without crossing each other. This is in agreement with the result
of Ref.~\cite{agkkw}. This implies that for couplings less than the critical
coupling, the lowest state is predominantly a one-boson configuration 
with its parton distribution exhibiting a peak at momentum fraction $x=1$. 

Our initial investigation beyond the critical coupling has  
revealed multiple level crossings at stronger couplings. As the coupling
steadily increases, the dominant peak of the parton distribution of the
lowest excitation shifts to lower $x$  values indicating the dominance of 
multi-parton configurations. Thus further investigations beyond the critical coupling
could reveal the detailed map of how a condensate of bosons steadily 
builds up. We plan to study this phenomena in detail and extrapolate to the
continuum if possible, despite the 
numerical challenges one faces  at strong coupling.  
In other words, can we enhance the details of kink formation with the higher resolution 
(higher $K$) capabilities achievable with modern supercomputers?

Can we return to the broken phase, which we have investigated previously with
both the PBC and APBC,  extract the critical coupling with 
enhanced precision and relate the critical couplings of the symmetric 
and the broken phases?  The weak/strong duality of this theory was discussed 
long ago by 
S.-J.~Chang~\cite{sjc}
and is studied recently in the equal-time Hamiltonian Fock space method
 in Ref.~\cite{rv2}. 

Can we make the connection with the results from equal-time calculations more 
precise following the insights provided by Refs.~\cite{bch,fkw}?
In this connection, a fundamental assumption is the equivalence between Light 
Front and Instant form field theories, which has been studied for a long time. 
Of primary concern is the fate of vacuum~\cite{weinberg,changma} and self energy (tadpole)
contributions. Here one analyzes Feynman diagrams expressed in light-cone
variables in order to clarify the potential importance of the zero 
longitudinal momentum mode. For more recent works see
Refs.~\cite{Mannheim:2020rod,Mannheim:2019lss,Polyzou:2021qpr}. 
Collins\cite{Collins:2018aqt} has given a
pedagogical treatment of the subject where a cure  is also provided for
circumventing the {\textquotedblleft}zero mode{\textquotedblright}
problem and for obtaining correct results in perturbation theory. 
For an application of this proposal within 
DLCQ see Ref.~\cite{md}.

Coming back to the matching of results from the Instant form and the Front form
theories, the problem arises mainly from the difference in mass corrections.
This leads to a difference in the effective dimensionless 
coupling $\lambda/\mu^2.$
 How to resolve this issue  non-perturbatively is presented
in Ref.~\cite{bch}.
A more ambitious program is presented in Ref.~\cite{fkw}. 
These corrections are addressable within DLCQ~\cite{Mengyao} 
which  require additional
calculations as defined in Ref.~\cite{fkw}. A detailed investigation of this matching using DLCQ
requires a separate future effort.

\section{Acknowledgements}
We thank John Hiller and Sophia Chabysheva for sharing details of their results.
We also thank Pieter Maris for useful discussions.
This work was supported in part by the US Department of
Energy (DOE) under Grant Nos.~DE-FG02-87ER40371 and DE-SC0018223 (SciDAC-4/NUCLEI). 
Computational resources were
provided by the National Energy Research Scientific Computing Center
(NERSC), which is supported by the US DOE Office of Science 
under Contract No. DE-AC02-05CH11231. 

%\newpage
%\section{Appendix}
%We present tables for DLCQ computational details and for the coefficients of the fit functions described in the text.

\end{document}